\documentclass[conference,a4paper]{APSIPA2021}
\usepackage{multirow}
\usepackage{graphicx}
\usepackage{amsmath}
\usepackage[psamsfonts]{amssymb}
\usepackage{amsxtra}
\usepackage{threeparttable}
\usepackage{booktabs}
\usepackage[labelformat=simple]{subcaption}

\begin{document}

\title{Efficient conformer-based speech recognition with linear attention}

\author{%
\authorblockN{%
Shengqiang Li, Menglong Xu, Xiao-Lei Zhang
}

\authorblockA{%
CIAIC, School of Marine Science and Technology, Northwestern Polytechnical University, China\\
E-mail: shengqiangli,mlxu@mail.nwpu.edu.cn, xiaolei.zhang@nwpu.edu.cn}
}

\maketitle
\thispagestyle{empty}

\begin{abstract}
Recently, conformer-based end-to-end automatic speech recognition, which outperforms recurrent neural network based ones, has received much attention. Although the parallel computing of conformer is more efficient than recurrent neural networks, the computational complexity of its dot-product self-attention is quadratic with respect to the length of the input feature. To reduce the computational complexity of the self-attention layer, we propose multi-head linear self-attention for the self-attention layer, which reduces its computational complexity to linear order. In addition, we propose to factorize the feed forward module of the conformer by low-rank matrix factorization, which successfully reduces the number of the parameters by approximate 50\% with little performance loss. The proposed model, named \textit{linear attention based conformer} (LAC), can be trained and inferenced jointly with the connectionist temporal classification objective, which further improves the performance of LAC. To evaluate the effectiveness of LAC, we conduct experiments on the AISHELL-1 and LibriSpeech corpora. Results show that the proposed LAC achieves better performance than 7 recently proposed speech recognition models, and is competitive with the state-of-the-art conformer. Meanwhile, the proposed LAC has a number of parameters of only 50\% over the conformer with faster training speed than the latter.
\end{abstract}

\section{Introduction}

End-to-end automatic speech recognition has received more and more attention in both academia and industry. It combines acoustic model, pronunciation model, and language model into a single neural network. It achieves competitive results with conventional speech recognition, and simplifies the training and
decoding pipelines. There are mainly three popular end-to-end approaches, named connectionist temporal classification (CTC) \cite{amodei2016deep}, recurrent neural network transducer \cite{graves2012sequence,sainath2020streaming}, and attention based encoder-decoder \cite{chan2016listen,vaswani2017attention}. This paper focuses on the attention based encoder-decoder model. There are two main kinds of attention based encoder-decoder models. The early models use recurrent neural networks as the building blocks of its encoder and decoder. Recently, a transformer architecture with a multi-head self-attention mechanism has become prevalent because of its high accuracy and fast training speed.

Although transformer is good at capturing content-based global interactions and can be trained in parallel, the computational complexity of its dot-product self-attention is quadratic with respect to the input feature length, which requires large GPU memory and long training time. Several studies have been conducted to reduce the computational complexity of the dot-product self-attention \cite{fujita2020attention,choromanski2020masked,choromanski2020rethinking,wang2020efficient,xu2020transformer,shen2021efficient,bello2021lambdanetworks}. In \cite{fujita2020attention}, Fujita \textit{et al.} replaced the self-attention layer with lightweight and dynamic convolution, where the convolution weight is dynamically predicted through an additional linear layer. In \cite{choromanski2020masked,choromanski2020rethinking}, choromanski \textit{et al.} introduced performer, which uses a novel fast attention via positive orthogonal random features approach (FAVOR+). Its complexity scales linearly rather than quadratically with respect to the number of the elements of an input sequence. Wang \textit{et al.} \cite{wang2020efficient} proposed to adopt the performer for efficiency improvement. In our previous work \cite{xu2020transformer}, we proposed local dense synthesizer attention which restricts the length of the attention weights for the reduction of the storage and computational complexity.

On the other hand, many works have shown that speech recognition models with small sizes can perform comparably to large models. For example, QuartzNet \cite{kriman2020quartznet}, which uses one-dimension separable convolution based models trained with the CTC loss, achieves the state-of-the-art accuracy on LibriSpeech. Winata \textit{et al.} \cite{winata2020lightweight} adopted a low-rank matrix factorization to compress models. The compressed models suffer little performance degradation. ContextNet \cite{han2020contextnet} combined a recurrent neural network transducer decoder with a fully convolutional encoder, where the encoder incorporates global context information into its convolution layers by squeeze-and-excitation modules.

In this paper, motivated by \cite{shen2021efficient} on the object detection task in video processing, we propose to apply the efficient attention \cite{shen2021efficient} to conformer and further extend it to multi-head version, named multi-head linear self-attention (MHLSA), for speech recognition. Specifically, MHLSA replaces the dot-product self-attention in the encoders of the conformer. It is functionally equivalent to the dot-product self-attention, but with substantially less computational costs than the latter, given that MHLSA does not need to calculate the correlation matrix in the dot-product self-attention.

To further reduce the model size of conformer, we factorize the feed forward module in both the encoder and decoder by low-rank matrix factorization. The overall model is named \textit{linear attention based conformer} (LAC). To improve the performance of LAC, we train and decode LAC jointly with the CTC objective. We conduct experiments on the AISHELL-1 \cite{bu2017aishell} and LibriSpeech \cite{panayotov2015librispeech} {corpora}. Experimental results show that the proposed LAC achieves a character error rate (CER) of $5.02\%$ on AISHELL-1, and a word error rate (WER) of $2.1\%$ and $2.3\%$ on the `dev-clean' and `test-other' sets of LibriSpeech respectively, which is competitive with the conformer \cite{guo2020recent}. Meanwhile, LAC has a number of parameters of only $50\%$ over the conformer, and accelerates the training speed by 1.23 times and 1.18 times on AISHELL-1 and LibriSpeech respectively.

The remainder of this paper is organized as follows. Section \ref{sec:model_description} describes the architecture of the proposed LAC. Section \ref{sec:experiment} presents experiments. Conclusion is given in Section \ref{sec:conclusion}.

\section{Linear attention based conformer} \label{sec:model_description}

\begin{figure}
	\centering
	\includegraphics[width=\linewidth]{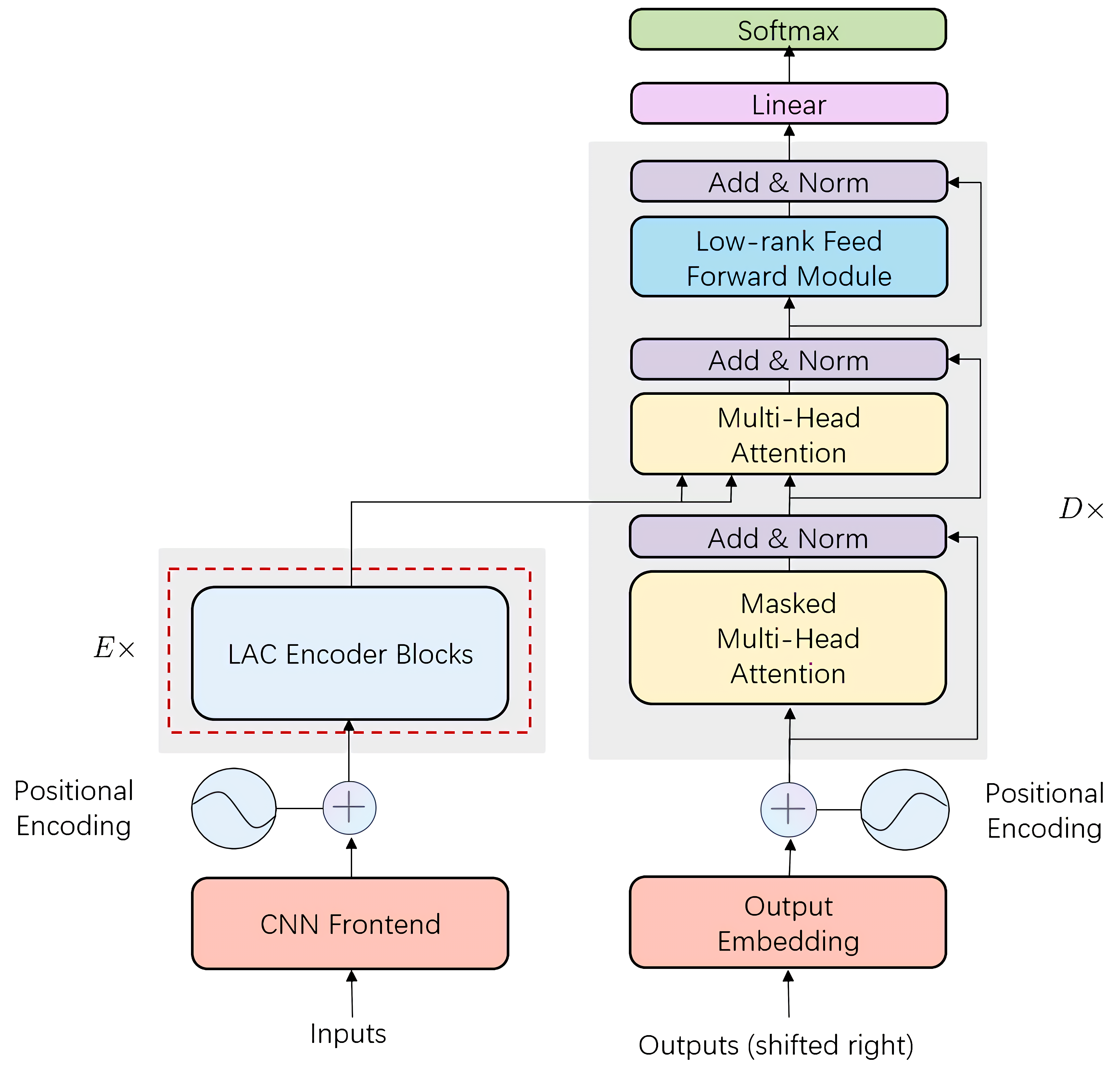}
	\caption{The model architecture of LAC.}
	\label{fig:lac}
\end{figure}

The architecture of the proposed LAC is shown in Figure \ref{fig:lac}. It includes a convolution neural network (CNN) frontend, a stack of encoders, and a stack of decoders. In this section, we focus on describing the CNN frontend in Section \ref{cnn frontend} and the encoder blocks of LAC in Section \ref{lac encoder}, as emphasized in the red dot box of Figure \ref{fig:lac}. Section \ref{complexity}  analyzes the complexity of the linear attention.

We will not describe the decoders of LAC anymore, given that the multi-head attention in the decoders has been described in detail in \cite{vaswani2017attention}, and also the low-rank feed forward module in the decoders is similar to the low-rank feed forward module in the encoders.

\subsection{CNN frontend}
\label{cnn frontend}
First, after passing through two CNN layers and an embedding layer, the input acoustic features of an utterance are transformed into a feature matrix $\mathbf{X}^{emb} \in \mathbb{R}^{T \times d_e}$, where $T$ is the sequence length after subsampling, and $d_e$ is the dimension of the embedding layer. Because our MHLSA does not need to compute the correlation matrix, we replace the relative position encoding in the conformer by absolute position encoding \cite{vaswani2017attention}. In this work, the sine and cosine functions are used for the absolute position encoding $\mathbf{P}\in \mathbb{R}^{T \times d_e}$, which is defined as:
\begin{equation}
	\begin{aligned}
		p_{i, 2 j} &=\sin \left( i / 10000^{2 j / d_e}\right) \\
		p_{i, 2 j+1} &=\cos \left( i / 10000^{2 j / d_e}\right)
	\end{aligned}
\end{equation}
where $i$ is the position, and $j$ is the dimension. Then, the position encoding $\mathbf{P}$ is added to the feature matrix $\mathbf{X}^{emb}$:
\begin{equation}	
	\mathbf{X}^{pos} = \mathbf{X}^{emb} + \mathbf{P}
\end{equation}
Finally, the matrix $\mathbf{X}^{pos}$ passes through a stack of LAC encoder blocks for the output of the encoder.

\subsection{Linear attention based conformer encoder}
\label{lac encoder}
\begin{figure}
	\centering
	\includegraphics[width=0.45\textwidth]{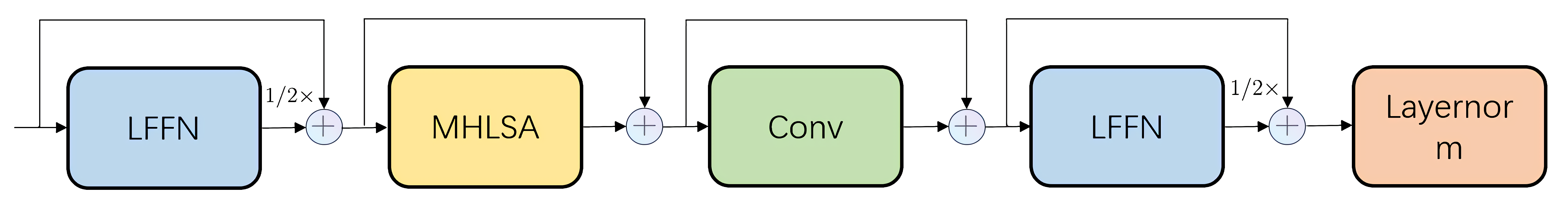}
	\caption{The encoder architecture of LAC.}
	\label{fig:lac_encdoer}
\end{figure}

The distinctive feature of our model is the use of LAC encoder blocks. Each LAC encoder block contains two low-rank feed forward (LFFN) modules sandwiching the MHLSA module and the convolution (Conv) module, as shown in Figure \ref{fig:lac_encdoer}.

Given an input $\mathbf{X}_{i}$ to the $i$-th encoder block, the output $\mathbf{Y}_i$ of the block is:
\begin{equation}
	\tilde{\mathbf{X}}_{i}=\mathbf{X}_{i}+\frac{1}{2} \operatorname{LFFN}\left(\mathbf{X}_{i}\right)
\end{equation}
\begin{equation}
	\mathbf{X}_{i}^{\prime}=\tilde{\mathbf{X}_{i}}+\operatorname{MHLSA}\left(\tilde{\mathbf{X}_{i}}\right)
\end{equation}
\begin{equation}
	\mathbf{X}_{i}^{\prime \prime}=\mathbf{X}_{i}^{\prime}+\operatorname{Conv}\left(\mathbf{X}_{i}^{\prime}\right)
\end{equation}
\begin{equation}
	\mathbf{Y}_{i}=\operatorname { Layernorm }\left(\mathbf{X}_{i}^{\prime \prime}+\frac{1}{2} \operatorname{LFFN}\left(\mathbf{X}_{i}^{\prime \prime}\right)\right)
\end{equation}
where $\operatorname{LFFN}(\cdot)$ refers to the low-rank feed forward module, $\operatorname{MHLSA}(\cdot)$ refers to the multi-head linear self-attention module, $\operatorname{Conv}(\cdot)$ refers to the convolution module, and $\operatorname{ Layernorm }$ refers to layer normalization. The convolution module is the same as that in \cite{gulati2020conformer}. It starts with a pointwise convolution layer and a gated linear unit, followed by a 1-D depthwise convolution layer. Batch normalization is applied after the convolution layers.
In the following subsections, we will introduce the proposed $\operatorname{LFFN}(\cdot)$ and $\operatorname{MHLSA}(\cdot)$ modules respectively.

\subsubsection{Low-rank feed forward module}
\label{LFFN}

Each of the layers in the encoder or decoder contains multiple feed forward modules. Each feed forward module consists of two linear transformations with an activation between them. Originally, the feed forward module is formulated as follows. Given an input $\mathbf{X}$ to the feed forward module, the output is:

\begin{equation}
		\operatorname{FFN}(\mathbf{X}) =  \operatorname{Dropout}(\operatorname{Swish}(\mathbf{X} \mathbf{W}_{1}))\mathbf{W}_{2}
\end{equation}
where $\operatorname{Swish}(\cdot)$ refers to the Swish activation, $\operatorname{Dropout}(\cdot)$ refers to the dropout operator, $\mathbf{W}_{1} \in \mathbb{R}^{d \times d_{ff}}$, $\mathbf{W}_{2} \in \mathbb{R}^{d_{ff} \times d}$ are the weight matrices of the two linear layers, $d$ and $d_{ff}$ denote the output dimension and hidden dimension of the feed forward module respectively.

\begin{figure}
	\centering
	\includegraphics[width=0.45\textwidth]{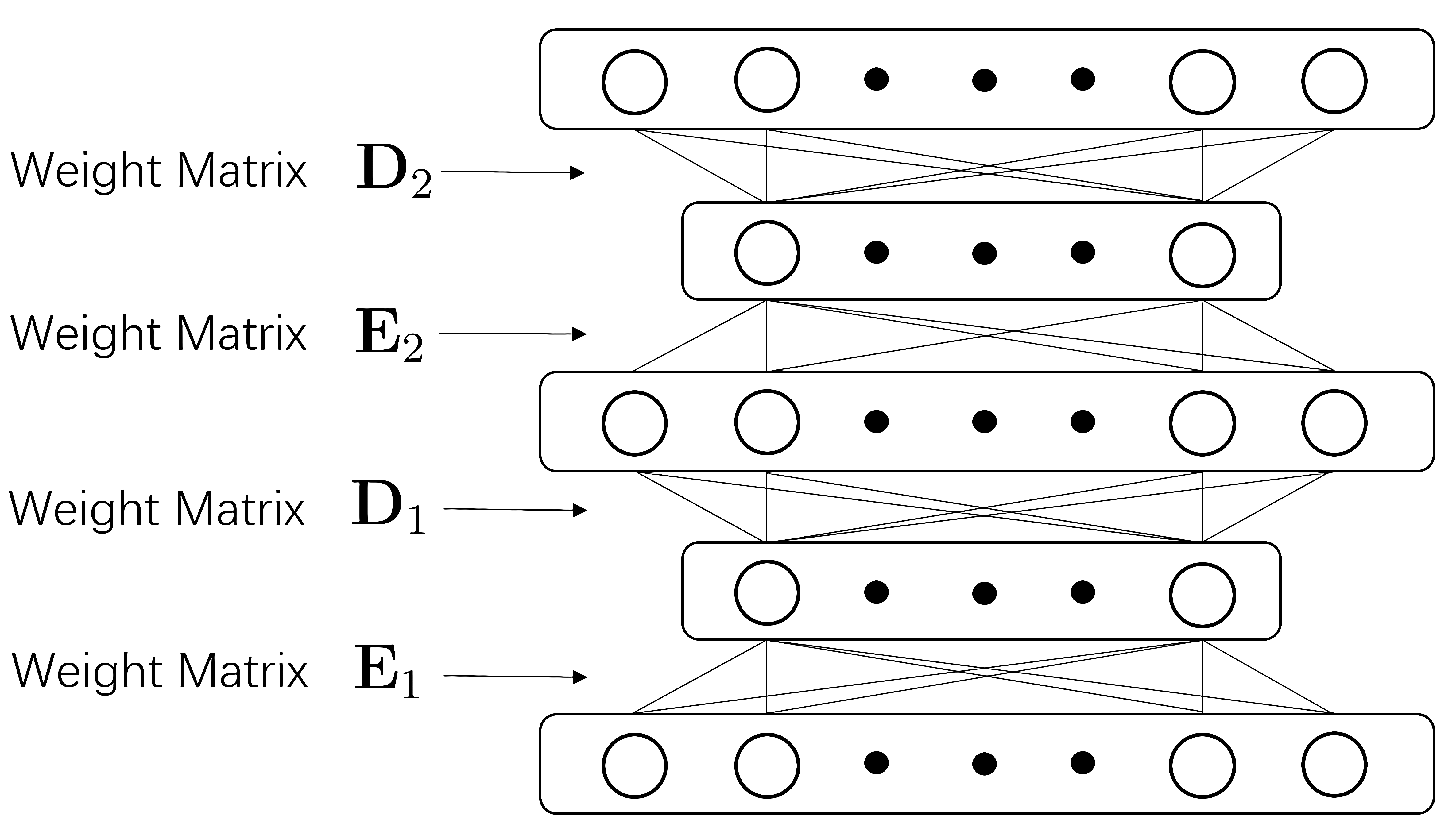}
	\caption{The low-rank feed forward module.}
	\label{fig:LFFN}
\end{figure}

To reduce the model size of the feed forward module, we factorize it by low-rank matrix factorization. Specifically, as shown in Figure \ref{fig:LFFN}, we factorize the weight matrix of the original linear layer $\mathbf{W}$ by two small matrices $\mathbf{E}$ and $\mathbf{D}$:

\begin{equation}
		\operatorname{LFFN}(\mathbf{X})=  \operatorname{Dropout}(\operatorname{Swish}(\mathbf{X} \mathbf{E}_{1} \mathbf{D}_{1}))\mathbf{E}_{2} \mathbf{D}_{2}
\end{equation}
where $\mathbf{E}_1 \in \mathbb{R}^{d \times d_{bn}}$ and $\mathbf{D}_1 \in \mathbb{R}^{d_{bn} \times d_{ff}}$ are the factorized matrices of $\mathbf{W}_1$, $\mathbf{E}_2 \in \mathbb{R}^{d_{ff} \times d_{bn}}$ and $\mathbf{D}_2 \in \mathbb{R}^{d_{bn} \times d}$ are the factorized matrices of $\mathbf{W}_2$, and $d_{bn}$ denotes the dimension of the bottleneck layer in the low-rank feed forward module.

Comparing the feed forward module with the low-rank feed forward module, we see that, the matrix $\mathbf{W}$ requires $d \times d_{ff}$ parameters and $d \times d_{ff}$ flops, while $\mathbf{E}$ and $\mathbf{D}$ only require $d_{bn} \times (d+d_{ff})$ parameters and flops. Because $d_{bn}$ is much smaller than $ d_{ff}$ and $d$, the number of the parameters and flops of $\mathbf{E}$ and $\mathbf{D}$ are much smaller than that of $\mathbf{W}$.

\subsubsection{Multi-head linear self-attention}
\label{MHLSA}

\begin{figure*}[t]
	\begin{subfigure}{0.36\linewidth}
		\centering
		\includegraphics[width=8.5cm]{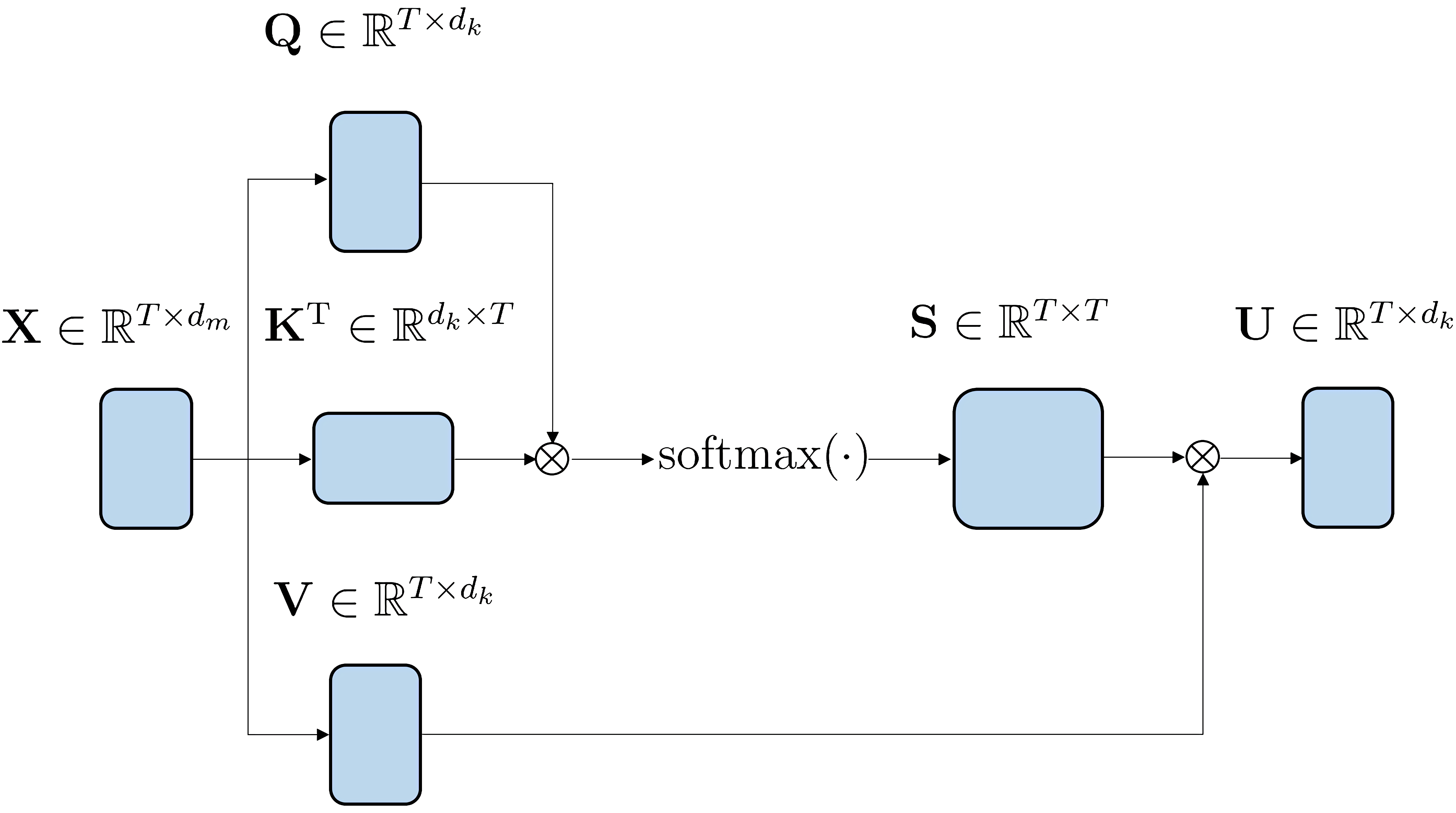}
		\caption{Dot-product self-attention.}
		\label{fig:attention_a}
	\end{subfigure}\hspace{-80mm}
	\hfill
	\begin{subfigure}{0.48\linewidth}
		\centering
		\includegraphics[width=8.5cm]{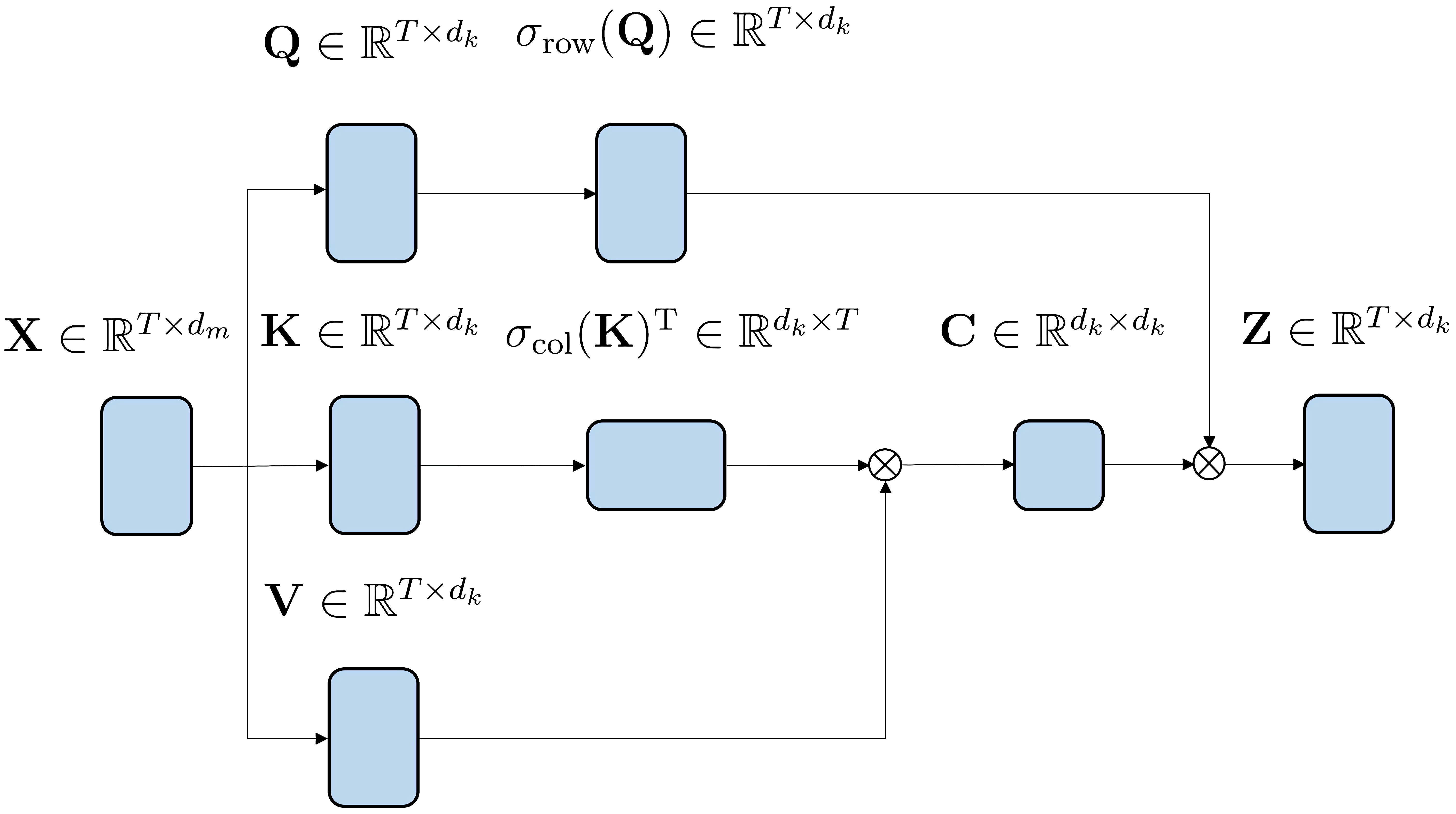}
		\caption{Linear self-attention}
		\label{fig:attention_b}
	\end{subfigure}
	\caption{Illustration of the architecture of dot-product self-attention and linear self-attention. $\text{T}$ denotes transpose operator, $\sigma_{\text{row}}$ and $\sigma_{\text{col}}$ refer to normalization functions operating in row and column of matrix $\mathbf{Y}$ respectively. }
	\label{fig:attentions}
\end{figure*}

We first revisit the multi-head dot-product self-attention in the conventional conformer as follows. For the $h$-th head of the multi-head dot-product self-attention, we pass the input $\mathbf{X} \in \mathbb{R}^{T \times d_m}$ through three linear projection layers which derives the query matrix $\mathbf{Q}_h \in \mathbb{R}^{T \times d_k}$, key matrix $\mathbf{K}_h \in \mathbb{R}^{T \times d_k}$, and value matrix $\mathbf{V}_h \in \mathbb{R}^{T \times d_k}$ by:
\begin{equation}
	\mathbf{Q}_{h}=\mathbf{X} \mathbf{W}^{\mathrm{Q}_{h}}, \mathbf{K}_{h}=\mathbf{X} \mathbf{W}^{\mathrm{K}_{h}}, \mathbf{V}_{h}=\mathbf{X} \mathbf{W}^{\mathrm{V}_{h}}
\end{equation}
where $\mathbf{W}^{\mathrm{Q}_{h}},\mathbf{W}^{\mathrm{K}_{h}},\mathbf{W}^{\mathrm{V}_{h}} \in \mathbb{R}^{d_m \times d_k} $ are the weight matrices of the three linear projection layers of the $h$-th head, $T$ is the length of the input feature, and $d_m$ is the hidden dimension of the multi-head self-attention layer. Assuming that there are $H$ heads in the multi-head self-attention layer, then $d_k=d_m/H$ is the dimension of each head. The output of the $h$-th dot-product self-attention module is
\begin{equation}
	\operatorname { Attention }(\mathbf{Q}_h, \mathbf{K}_h, \mathbf{V}_h)=\operatorname{softmax}\left(\frac{\mathbf{Q}_h \mathbf{K}_{h}^{\text{T}}}{\sqrt{d_k}}\right) \mathbf{V}_h
	\label{dot-product attention_1}
\end{equation}
The dot-product self-attention based multi-head self-attention is formulated as:
\begin{equation}
	\operatorname{MHSA}(\mathbf{Q}, \mathbf{K}, \mathbf{V})=\operatorname { Concat }\left(\mathbf{U}_{1}, \cdots, \mathbf{U}_{H}\right) \mathbf{W}^{\mathrm{O}}
\end{equation}
where
\begin{equation}
	\mathbf{U}_{h}=\operatorname { Attention }\left(\mathbf{X} \mathbf{W}^{\mathrm{Q}_{h}}, \mathbf{X} \mathbf{W}^{\mathrm{K}_{h}}, \mathbf{X} \mathbf{W}^{\mathrm{V}_{h}}\right)
\end{equation}
and $\mathbf{W}^{\mathrm{O}} \in \mathbb{R}^{d_m \times d_m}$ is the parameter matrix of the output projection layer.

To reduce the computational complexity of the self-attention layer to linear order, we first replace the dot-product self-attention by linear (self-)attention, and then extends the linear attention to MHLSA. Specifically, for the $h$-th head in MHLSA, the input $\mathbf{X} \in \mathbb{R}^{T \times d_m}$ is transformed to
the query matrix $\mathbf{Q}_h \in \mathbb{R}^{T \times d_k}$, key matrix $\mathbf{K}_h \in \mathbb{R}^{T \times d_k}$, and value matrix $\mathbf{V}_h \in \mathbb{R}^{T \times d_k}$ by three linear projection layers respectively. The output of the $h$-th head MHLSA is
\begin{equation}
	\operatorname{LinearAtt}(\mathbf{Q}_h, \mathbf{K}_h, \mathbf{V}_h)=\sigma_{\text{row}}\left(\frac{\mathbf{Q}_h}{{d_k}^{\frac{1}{4}}}\right)\left(\sigma_{\text{col}}\left(\frac{\mathbf{K}_h}{{d_k}^{\frac{1}{4}}}\right)^{\text{T}} \mathbf{V}_h\right)
\end{equation}
where $\sigma_{\text{row}}(\cdot)$ and $\sigma_{\text{col}}(\cdot)$ denote the operators of applying the softmax function along the rows and columns of a matrix respectively, and $^{\text{T}}$ denotes the transpose operator. Finally, MHLSA is formulated as:
\begin{equation}
	\operatorname{MHLSA}(\mathbf{Q}, \mathbf{K}, \mathbf{V})=\operatorname { Concat }\left(\mathbf{Z}_{1}, \cdots, \mathbf{Z}_{H}\right) \mathbf{W}^{\mathrm{O}}
\end{equation}
where
\begin{equation}
	\mathbf{Z}_{h}=\operatorname{LinearAtt}(\mathbf{Q}_h, \mathbf{K}_h, \mathbf{V}_h).
\end{equation}

\subsection{Computational complexity analysis}
\label{complexity}
As shown in Figure \ref{fig:attention_a}, because the dot-product self-attention needs to calculate an attention weight matrix $ \mathbf{S} \in \mathbb{R}^{T \times T} $, the computational complexity of each head is as high as $O\left(T^{2} d_k \right)$.

As a comparison, the linear attention, which is described in Figure \ref{fig:attention_b}, switches the matrix multiplication order from $\left(\mathbf{Q K}^{\text{T}}\right) \mathbf{V}$ to $\mathbf{Q}\left(\mathbf{K}^{\text{T}}\mathbf{V}\right) $, which reduces the computational complexities of each head from $O\left(T^{2} d_k \right)$ to $O\left(T d_{k}^2 \right)$ without affecting the effectiveness of the matrix multiplication. One can see that the linear attention is mathematically equivalent to the dot-product self-attention, however, it is substantially faster and needs less memory than the latter.

\section{Experiments} \label{sec:experiment}

\subsection{Experiment Setup}

Our experiments were conducted on a Mandarin speech corpus AISHELL-1 \cite{bu2017aishell} and an English speech corpus LibriSpeech \cite{panayotov2015librispeech}. The former has 170 hours labeled speech, while the latter consists of 970 hours labeled speech and an additional 800M word token text-only corpus for building language model.

We used 80-channel log-mel filterbank coefficients (Fbank) features computed on a 25ms window with a 10ms shift. The features for each speaker were rescaled to have zero mean and unit variance. The token vocabulary of AISHELL-1 contains 4231 characters. We used a 5000 token vocabulary based on the byte pair encoding algorithm \cite{sennrich2015neural} for LibriSpeech. Moreover, the vocabularies of AISHELL-1 and LibriSpeech have a padding symbol '$\langle PAD\rangle$' , an unknown symbol '$\langle UNK\rangle$', and an end-of-sentence symbol '$\langle EOS\rangle$'.

Our LAC model contains 12 encoder blocks and 6 decoder blocks. There are 4 heads in both the multi-head linear self-attention and the encoder-decoder attention. The 2D-CNN frontend utilizes two $3\times3$ convolution layers with 256 channels. The rectified linear units were used as the activation. The stride was set to 2. The hidden dimension of the attention layer is 256. The hidden dimension and output dimension of the feed forward layer are 256 and 2048 respectively. We used the Adam optimizer and a transformer learning rate schedule \cite{vaswani2017attention} with 30000 warm-up steps and a peak learning rate of 0.0005. We used SpecAugment \cite{park2019specaugment} for data augmentation. We set the CTC weight to 0.3 for the joint training with the attention model. In the test stage, we set CTC weight to 0.6 for the joint decoding. We used a transformer-based language model to refine the results.

We choose 7 representative transformer-based speech recognition models as the comparison methods so that we can fairly compare our model with other state-of-the-art models in the computational complexity of the self-attention, number of parameters, CER and WER. The comparison methods are the speech transformer (ST) \cite{tian2020spike}, local dense synthesizer attention (LDSA) \cite{xu2020transformer}, low-rank transformer (LRT) \cite{winata2020lightweight}, dynamic convolution (DC) \cite{fujita2020attention}, self-attention with 2D dynamic convolution (SA-DC2D) \cite{fujita2020attention}, performers in conformer (PIC) \cite{wang2020efficient}, and conformer \cite{guo2020recent}. Speech transformer adopts the transformer architecture as its encoder network and prediction network. LRT applies matrix decomposition to the transformer architecture. LDSA replaces the self-attention in transformer encoder with the local dense synthesizer attention. PIC adopts the performer to improve the efficiency of conformer. DC and SA-DC2D apply dynamic convolution to transformer as an alternative architecture to the self-attention module. Conformer combines convolution neural networks and transformers to model both local and global dependencies. To compare fairly with conformer, we implemented an absolute position encoding based conformer whose parameter settings are the same with LAC except the two novel points, i.e. the attention module in the encoder and the low-rank feed forward module.

\begin{table}[t]
	\caption{Comparison results on AISHELL-1. The symbol `$T$' represents the length of input feature. The symbol `$c$' in LDSA denotes the context width. The term `\# Param.' is short for the number of parameters.}
	\vspace{6pt}
	\label{tab:compare_b}
	\centering
	\scalebox{0.85}{
		\begin{tabular}{lllll}
			\toprule
			
			\multirow{2}{*}{\textbf{Model}} &  \multirow{2}{*}{\textbf{Complexity}} & \multirow{2}{*}{\textbf{\# Param.}} & \multicolumn{2}{c}{\textbf{CER (\%)}} \\
			& & & \textbf{Dev set} & \textbf{Test set} \\
			\midrule
			
			ST \cite{tian2020spike} & $O(T^2)$ & 33.91M & 6.57 & 7.37 \\
			
			LRT \cite{winata2020lightweight} & $O(T^2)$ & 12.70M & - & 13.09 \\
			
			LDSA \cite{xu2020transformer} & $O(Tc)$ & 52.5M & 5.79 & 6.49 \\
			
			Conformer \cite{guo2020recent} & $O(T^2)$ & 45.15M  & 4.52 & 4.88 \\
			
			LAC & $O(T)$ & 22.83M  & 4.73 & 5.02 \\				
			
			\bottomrule
			
	\end{tabular}}
\end{table}

\begin{table}[t]
	\caption{Comparison results on LibriSpeech.}
	\vspace{6pt}
	\label{tab:compare_c}
	\centering
	\scalebox{0.85}{
		\begin{tabular}{lllllll}
			\toprule
			
			\multirow{3}{*}{\textbf{Model}} &  \multirow{3}{*}{\textbf{Complexity}} & \multirow{3}{*}{\textbf{\# Param.}} & \multicolumn{4}{c}{\textbf{WER(\%)}} \\
			& & & \multicolumn{2}{c}{\textbf{Dev}} & \multicolumn{2}{c}{\textbf{Test}} \\
			& & & \textbf{Clean} & \textbf{Other} & \textbf{Clean} & \textbf{Other} \\
			\midrule

			DC \cite{fujita2020attention} & $O(T)$ & - & 3.5 & 10.5 & 3.6 & 10.8 \\
			
			SA-DC2D \cite{fujita2020attention} & $O(T^2)$ & - & 3.5 & 9.6 & 3.9 & 9.6 \\
			
			PIC \cite{wang2020efficient} & $O(T)$ & 10.4M & 2.9 & 7.2 & 3.1 & 7.7 \\
			
			Conformer \cite{guo2020recent} & $O(T^2)$ & 45.16M  & 2.1 & 5.5 & 2.3 & 5.5 \\
			
			LAC & $O(T)$ & 23.47M  & 2.1 & 5.6 & 2.3 & 5.8 \\				
			
			\bottomrule
			
	\end{tabular}}
\end{table}

\begin{table}[t]
	\caption{Effect of the bottleneck dimension $d_{bn}$ of LAC on performance.}
	\vspace{6pt}
	\label{tab:compare_a}
	\centering
	\scalebox{0.85}{
		\begin{tabular}{lllll}
			\toprule
			
			\multirow{2}{*}{\textbf{Model}} &  \multirow{2}{*}{\textbf{Complexity}} & \multirow{2}{*}{\textbf{\# Param.}} & \multicolumn{2}{c}{\textbf{CER}} \\
			&  & & \textbf{Dev Set} & \textbf{Test Set} \\
			\midrule
			
			LAC ($d_{bn}=50$) & $O(T)$ & 17.07M  & 4.85 & 5.16 \\
			
			LAC ($d_{bn}=75$) &  $O(T)$ & 19.95M & 4.81 & 5.12 \\
			
			LAC ($d_{bn}=100$) & $O(T)$ & 22.83M & 4.73 & 5.02 \\
			
			LAC ($d_{bn}=125$) & $O(T)$ & 25.71M & 4.71 & 5.02\\
			
			\bottomrule
	\end{tabular}}
\end{table}

\begin{table}[t]
	\caption{Disentangling LAC. We replaced the core modules of the LAC encoder with the counterpart of conformer by (i) replacing low-rank feed forward (LFFN) module in LAC with feed forward (FFN) module in conformer, and (ii) replacing the MHLSA in LAC with the MHSA in conformer.}
	\vspace{6pt}
	\label{tab:ablation_study}
	\centering
	\scalebox{0.85}{
		\begin{tabular}{lllll}
			\toprule
			
			\multirow{2}{*}{\textbf{Model}} &  \multirow{2}{*}{\textbf{Complexity}} & \multirow{2}{*}{\textbf{\# Param.}} & \multicolumn{2}{c}{\textbf{CER}} \\
			&  & & \textbf{Dev set} & \textbf{Test set} \\
			\midrule
			
			LAC & $O(T)$ & 22.83M  & 4.73 & 5.02 \\
			
			$-$LFFN$+$FFN &  $O(T)$ & 45.10M & 4.54 & 4.90 \\
			
			$-$MHESA$+$MHSA & $O(T^2)$ & 22.83M & 4.67 & 4.99 \\
			
			Conformer \cite{guo2020recent} & $O(T^2)$ & 45.15M & 4.52 & 4.88 \\
			
			\bottomrule
			
	\end{tabular}}
\end{table}

\subsection{Main results}

Table \ref{tab:compare_b} lists the comparison result on AISHELL-1. From the table, we see that the proposed LAC achieves a CER of $4.73\%$ on the development set and $5.02\%$ on the test set, which is slightly worse than conformer {\cite{guo2020recent}}. The main advantage of LAC is that its parameter number and computational complexity are significantly lower than conformer, with a training speed of 1.23 times faster than the latter. Moreover, LAC significantly outperforms the other comparison methods.

Table \ref{tab:compare_c} lists the comparison result on the LibriSpeech dataset. From the table, we see that the proposed LAC achieves a competitive result with conformer \cite{guo2020recent}, and outperforms the other comparison methods. Although PIC has the smallest model size and a low computational complexity, its performance is not as good as LAC.

\subsection{Effect of the bottleneck dimension $d_{bn}$ of LAC.}

We studied the effect of the bottleneck dimension $d_{bn}$ of LAC on AISHELL-1. The experimental result in Table \ref{tab:compare_a} shows that setting $d_{bn}=50$ achieves the smallest model size with worst performance and setting $d_{bn}=100$ yields a good trade-off between the model size and the performance.

\subsection{Ablation studies}

The proposed LAC differs from conformer in the multi-head linear self-attention and low-rank feed forward module. Table \ref{tab:ablation_study} shows the effect of each change to conformer. Low-rank feed forward module reduces the number of parameters by approximate 50\% with little performance degradation. Multi-head linear self-attention reduces the computational complexity of encoder from $O(T^2)$ to $O(T)$ without performance degradation. The proposed LAC achieves a good trade-off among model size, computational complexity, and performance.

\section{Conclusions} \label{sec:conclusion}

In this paper, we have proposed the acoustic model ``linear attention based conformer'' for speech recognition. Specifically, we first extended the linear attention originally proposed for object detection to a multi-head version and then applied it to speech recognition. It reduces the computational complexity of the self-attention layers in the conformer encoder to a linear order with respect to the input feature length. In addition, we also applied the low-rank matrix factorization to the feed forward module in conformer, which reduces the number of the parameters of conformer by approximate 50\%. To improve the performance of LAC, we train and decode it jointly with the CTC objective. Experimental results on AISHELL-1 and LirbriSpeech show that the proposed LAC achieves competitive performance with conformer, and performs better than 7 other recent speech recognition models.

\bibliographystyle{IEEEtran}
\bibliography{refs}

\end{document}